# The Random Buffer Tree: A Randomized Technique for I/O-efficient Algorithms

Saju Jude Dominic [1], G. Sajith[1]

[1] Indian Institute of Technology, Guwahati, India.
{sjdom,sajith}@iitg.ernet.in

**Abstract.** In this paper, we present a probabilistic self-balancing dictionary data structure for massive data sets, and prove expected amortized I/O-optimal bounds on the dictionary operations. We show how to use the structure as an I/O-optimal priority queue. The data structure, which we call as the random buffer tree, abstracts the properties of the random treap and the buffer tree. The random buffer tree immediately yields an I/O-optimal randomized algorithm for sorting numbers in external memory. We discuss heuristics to transform the random buffer tree into a self-adjusting data structure suitable for applications with skewed access patterns.

## 1 Introduction

The study of I/O-efficient algorithms and data structures has been receiving increased attention in recent years due to the fact that communication between fast internal memory and slower external memory such as disks is the bottleneck in many computations involving massive datasets. The significance of this bottleneck is increasing as internal computation gets faster and especially as parallel computing gains in popularity.

In this paper, we continue the development of I/O-efficient data structures, and illustrate further how they can assist in the design of I/O-efficient algorithms. The data structure we present is called the random buffer tree and abstracts the properties of the random treap [8] and buffer tree [4]. We show that the random buffer tree supports insert, delete, search and deletemin operations within an expected amortized optimal number of I/O's. The random buffer tree immediately yields a simple I/O-optimal randomized algorithm for sorting numbers in external memory using the standard tree-sort.

We provide a comparative study of the random buffer tree and the buffer tree and point out the advantages and disadvantages of each structure. Finally we explore some heuristics to transform the random buffer tree into a self-adjusting index suitable for applications with skewed access frequencies.

### 1.1 Model of Computation

We will be working in an I/O model introduced by Aggarwal and Vitter [1]. Our computational model consists of a single processor with a small local memory connected to a large external memory, which may consist of multiple parallel disks. The parameters of the model are as follows
- N = the total size of the problem in external memory,
- M = the number of items that can fit into the local memory,
- B = the number of items in one block of data,

It is assumed that M < N and 1 << B < M/2 . The model captures the essential parameters of many of the I/O-systems in use today, and depending on the size of the data elements, typical values are of the order of M = $10^6$ or $10^7$ and B = $10^3$. Large scale problem instances can be in the range N = $10^{10}$ or $10^{12}$.

An I/O operation in the model is a swap of *B* elements from internal memory with *B* consecutive elements from external memory. The measure of performance we consider is the number of such I/O's needed to solve a given problem. The quotients N/B (the number of blocks in the problem) and M/B (the number of blocks that fit into internal memory) play an important role in the study of I/O-complexity. Therefore we will use *n* as shorthand for N/B and *m* for M/B.

### 1.2 Previous Results

Aggarwal and Vitter [1] first considered the problem of designing I/O-efficient algorithms. They gave several algorithms for basic problems such as sorting, permuting, and matrix operations. Goodrich et al. [2] and Chiang et al. [3] developed external memory techniques applicable to computational geometry and graph problems respectively. Arge [4] improved several of the results in these earlier papers by the introduction of the I/O efficient *buffer tree*, which was the first I/O-efficient data structure to incorporate an amortized analysis for batched operations. Kumar et al. [5] introduced the I/O-efficient heap based on the buffer tree.

### 1.3 Our results

In this paper, we continue the development of I/O-efficient offline data structures, and illustrate further how they can assist in the design of I/O-efficient algorithms. The data structure we present is called as the *random buffer tree* and it abstracts the properties of the random treap and the buffer tree. We show that the random buffer tree supports insert, delete, search and deletemin operations in an expected amortized bound of O ($\frac{\log_m n}{B}$) I/O's per operation which is optimal. The random buffer tree has the same expected performance bounds as Arge's buffer tree in the worst case and has a considerably simpler specification compared to the buffer tree.

The remainder of the paper is organized as follows. We present the random buffer tree in Section 2.2 and describe the primitives and operations allowed on the structure in Section 2.3 and 2.4 respectively. We establish expected amortized I/O-optimal bounds on the operations in Section 25. We compare random buffer tree with the buffer tree in Section 2.6 and discuss the heuristics to make the random buffer tree self-adjusting in Section 2.7 . Finally, we state our conclusions in Section 3.

## 2 The Random Buffer Tree

In this Section, we present our specification for the random buffer tree, a probabilistic self-balancing data structure for external memory offline applications, having the same expected amortized performance bounds as the buffer tree in the worst case.

### 2.1 The Basic Idea

The random buffer tree is a randomized $m$-ary search tree in which the priorities are independently and uniformly distributed continuous random variables. The N elements in the random-buffer tree are arranged in in-order with respect to keys and in (max-) heap order with respect to priorities. Whenever we insert a new element into the random buffer tree, we generate a random real number between (say) 0 and 1 and use that number as the priority of the new element. The only reason we're using real numbers is so that the probability of two nodes having the same priority is zero, since equal priorities make the analysis messy. In practice, we could just choose random integers from a large range, like 0 to $2^{31}$ - 1, or random floating point numbers. Also, since the priorities are independent, each node is equally likely to have the largest priority. The cost of all the operations on the random buffer tree - search, insert, delete and deletemin is proportional to the height of the tree. Devroye [6] has shown that the expected height of a random $m$-ary search tree is $O(\log_m n)$ and we show that that the expected running time for any of the above operations is $O(\frac{\log_m n}{B})$ .

The operations on the structure, insert, delete as well as queries, are done in a lazy manner as in the buffer tree. If we, for example, want to insert an element in the tree, we do not search down the tree to find the place among the leaves to insert the element. Instead, we wait until we have collected a block of insertions (and/or other operations), and then we insert this block in the buffer of the root node. When the buffer becomes full, we apply the updates and push all but the largest $m/2$ blocks of elements with respect to the priority values, one level down the tree, in search order of the key values, to the buffers on the next level. Similarly, when the buffer has less than $m/4$ blocks of elements as a result of deletions, we pick enough largest elements with respect to priority from among the child buffers to raise the number of elements in the buffer to $m/2$ blocks. We call these the buffer-emptying and buffer-filling processes respectively. Updates, as well as queries, are basically done in the same way as insertions. It means that queries get batched in the sense that the result of a query may be generated (and reported) lazily by several buffer-emptying processes.

The main requirement needed to satisfy the I/O bound of $O(\frac{\log_m n}{B})$ is that we should be able to empty a buffer in a number of I/O's that is linear in the number of elements in the buffer. If this is the case, we can do an amortization argument by associating a number of credits to each block of elements in the tree. More precisely, each block in the buffer of node $x$ must hold O(the height of the tree rooted at $x$) credits. As we only empty a buffer when it becomes full, the blocks in the buffer can pay for the emptying process as they all get pushed one level down. We give each update or query element $O(\frac{\log_m n}{B})$ credits on insertion in the root node, and this gives us the desired bounds.

### 2.2 Description of the data structure

The random buffer tree is a random $m$-ary search tree with the following properties:
- Elements have a *key* field and a *priority* field, which is a random number chosen independently for each element from a uniform continuous distribution.
- Each node has $m$ children.
- Each node has a *buffer* of size $M$ associated with it, which contains upto $m$ blocks of elements *stored in sorted order w.r.t. priority values.*
- Any element held in a node's buffer has a larger *priority* value than any other element held in the buffers of any of its child nodes.
- Any element held in a node's buffer has a larger *key* value than any other element stored in the buffers of its left siblings and a smaller *key* value than any other element stored in the buffers of its right sibling.
- Every internal node of the tree holds $\Theta(m)$ blocks of elements.
- The tree defines a $m$-ary max heap over the elements with respect to priority values and a m-ary search tree with respect to the key values.

We shall now state a result on the expected height of a random $m$-way search tree, due to Devroye [6].

Lemma 1: The expected height of a random $m$-way search tree over $n$ elements is $O(\log_m n)$.

*Proof*: A random $m$-ary search tree is constructed from a random permutation of 1, 2,…,n. A law of large numbers is obtained for the height $H_n$ of these trees by applying the theory of branching random walks. In particular, it is shown that $H_n/\log n \to g$ in probability as $n \to \infty$, where $g=g(m)$ is a constant depending upon m only. Interestingly, as $m \to \infty$, $g(m)$ is asymptotic to $1/\log m$, the coefficient of $\log n$ in the asymptotic expression for the height of the complete $m$-ary search tree. This proves that for large $m$, random $m$-ary search trees behave virtually like complete $m$-ary trees. See Devroye [6] for the complete proof.

### 2.3 The Random Buffer Tree Primitives

The standard operations *insert*, *delete*, and *deletemin* are supported, and are implemented using the following primitives: EMPTYBUFFER and FILLBUFFER.

*2.3.1 EMPTYBUFFER*
EMPTYBUFFER is executed whenever a node's list has over $m$ blocks of elements. This primitive is used to empty all but the largest $m/2$ blocks of elements with respect to the priority values, into the buffers at the next level in search order of the key values. The steps involved in EMPTYBUFFER are:
1. The $m$ blocks of elements are loaded into main memory in $O(m)$ I/O's which are then sorted w.r.t. the priority values. This step requires $\Theta(m)$ I/O's as $m$ blocks of elements are loaded into main memory.
2. In case there are delete requests, delete the requested element as well as the delete request element. This step requires no I/O's as all computations are done on elements in main memory.
3. If the buffer now contains more than $(3/4)m$ blocks of elements, all but the largest $m/2$ blocks of elements with respect to priority are pushed to the buffers at the next lower level in search order of the key values. The distribution is done such that the number of elements in each child buffer is as evenly balanced as possible. This step requires $\Theta(m)$ I/O's as $O(m)$ blocks of elements are moved to $O(m)$ buffers at the next level.
4. If the number of elements in the node's buffer now falls below $m/4$ blocks as a result of Step 2, FILLBUFFER is executed on the node's buffer to bring the number of elements in the buffer to $m/2$ blocks. This step requires $\Theta(m)$ I/O's as explained in Section 2.3.2.
5. If any of the children's buffer now contains more than $m$ blocks of elements, EMPTYBUFFER is recursively executed on the child buffer.

We will now state with proof some simple lemmas about EMPTYBUFFER.

**Lemma 2.1**: The cost of EMPTYBUFFER is $\Theta(m)$ I/O's.
*Proof*: Step 1 requires $\Theta(m)$ I/O's as $m$ blocks of elements are loaded into main memory. Step 2 requires no I/O's as all computations are done on elements in main memory. Step 3 requires $O(m)$ I/O's as at most $m/2$ blocks of elements are moved to at most $m$ buffers at the next level. Steps 4 requires $\Theta(m)$ I/O's as discussed in Section 2.3.2. Step 5 is a recursive call to the buffer-emptying routine. The result follows.

**Lemma 2.2**: EMPTYBUFFER maintains the invariant that each node holds $\Theta(m)$ blocks of elements.
*Proof*: The number of elements in the buffer before Step 1 is $m$ blocks. At the end of Step 2 the buffer has between zero to $m$ blocks of elements. If the number of elements is below $m/4$ blocks Step 4 is executed after which the number of elements in the buffer is $m/2$ blocks as explained in Section 2.3.2. Otherwise, if the number of elements is above *(3/4)m* blocks, Step 3 is executed after which the number of elements in the buffer is *m/2* blocks. Thus at the end of EMPTYBUFFER, the buffer has between *m/2* and *(3/4)m* blocks of elements. The result follows.

**Lemma 2.3**: The next EMPTYBUFFER on any node is guaranteed not to occur for the next $\Omega(m)$ blocks of updates (or other operations).

*Proof*: This is a simple consequence of Lemma 2.2 . Since at the end of EMPTYBUFFER, the buffer has between *m/2* and *(3/4)m* blocks of elements and EMPTYBUFFER is executed only when a node's buffer has over *m* blocks of elements, the next EMPTYBUFFER on the buffer does not occur for the next *m/4* to *m/2* updates(or other operations). The result follows.

**Lemma 2.4**: EMPTYBUFFER maintains the invariant that any element held in a node's buffer has a larger priority value than any other element held in the buffers of any of its child nodes.

*Proof*: Assume that the invariant is true at the beginning of Step 1 of EMPTYBUFFER. Steps 1 and 2 do not affect the invariant property. Step 3 maintains the invariant as the larger elements, with respect to priority, are retained in the buffer and lighter elements, with respect to priority, are pushed one level down the tree. Step 4 maintains the invariant as explained in Section 2.3.2. Hence if the invariant holds at the beginning of EMPTYBUFFER, then it also holds at the end of EMPTYBUFFER. The base case of empty tree is trivial. The result follows.

**Lemma 2.5**: EMPTYBUFFER maintains the invariant that any element held in a node's buffer has a larger key value than any other element stored in the buffers of its left siblings and a smaller key value than any other element stored in the buffers of its right sibling.

*Proof*: This is guaranteed by the manner in which the elements in the list are distributed to the buffers of the children(in search order of the key values) in Step 3 of EMPTYBUFFER. The other steps do not affect the invariant property. The result follows.

*2.3.2 FILLBUFFER*

FILLBUFFER is used to fill up a node's buffer when the number of elements in a node's buffer drops below *m/4* blocks. The FILLBUFFER primitive ensures that every internal node contains $\Omega(m)$ blocks of elements.

The steps involved in FILLBUFFER are:

1. Empty the buffer completely i.e. perform a full EMPTYBUFFER on the node's buffer. This step requires $O(m)$ I/O's as explained in Section 2.3.1.
2. Extract enough largest elements , with respect to priority values (up to *m/2* blocks) from the child buffers to raise the number of elements in the buffer to *m/2* blocks, by treating the child buffers as a collection of lists sorted with respect to priority values. This step requires $O(m)$ I/O's as upto *m/2* blocks of elements are moved from the child buffers to the parent buffer.

3. If any of the child buffers now has less than *m/4* blocks of elements, FILLBUFFER is recursively executed on the child buffer.

We will now state with proof some simple lemmas about FILLBUFFER.

**Lemma 3.1**: The cost of FILLBUFFER is $\Theta(m)$ I/O's.
*Proof*: Step 1 requires $O(m)$ I/O's as up to m/4 blocks of elements are emptied to the child buffers. Step 2 requires $O(m)$ I/O's as upto m/2 blocks of elements are moved from the child buffers to the parent buffer. The result follows.

**Lemma 3.2**: FILLBUFFER maintains the invariant that each node holds $\Theta(m)$ blocks of elements.
*Proof*: The number of elements in the buffer before Step 1 of FILLBUFFER is less than *m/4* blocks, which violates the invariant. At the end of Step 2 the list has *m/2* blocks of elements and the invariant is re-established. The result follows.

**Lemma 3.3**: The next FILLBUFFER on the node is guaranteed not to occur for the next $\Omega(m)$ blocks of updates (or other operations).
*Proof*: This is a simple consequence of Lemma 2.3. Since Lemma 2.3 guarantees that the next EMPTYBUFFER does not occur for the next $\Omega(m)$ blocks of updates (or other operations) and since FILLBUFFER is triggered only as a result of EMPTYBUFFER on a node, at least $\Omega(m)$ blocks of updates (or other operations) should occur between the last FILLBUFFER on the node and the next one. The result follows.

**Lemma 3.4**: FILLBUFFER maintains the invariant that any element held in a node's buffer has a larger key value than any other element held in the buffers of any of its child nodes.
*Proof*: Assume that the invariant is true at the beginning of Step 1 of FILLBUFFER. Step 2 restores the invariant property as the largest *m/2* blocks of elements, with respect to the priority, from among the child buffer are moved up to the parent buffer. Hence if the invariant holds at the beginning of FILLBUFFER, then it also holds at the end of FILLBUFFER. The result follows.

**Lemma 3.5**: FILLBUFFER maintains the invariant that any element held in a node's buffer has a larger key value than any other element stored in the buffers of its left siblings and a lesser key value than any other element stored in the buffers of its right sibling.
*Proof*: Assume that the invariant is true at the beginning of Step 1 of FILLBUFFER. Step 1 does not violate the invariant property as EMPTYBUFFER maintains the invariant by Lemma 2.5. Step 2 maintains the invariant. The result follows.

### 2.4 Random Buffer Tree Operations

The random buffer tree operations *insert*, *delete*, *search* and *deletemin* are described next.

*2.4.1 insert(key)*
The insert operation is used to insert a new element with the given *key*, into the tree. The steps involved in *insert* are:
1. Construct a new (insert) element consisting of the new element to be inserted, having the given key and assign a random priority from a uniform and continuous random distribution.
2. When *B* such insert elements have been collected in internal memory, insert the block in the buffer of the root node.
3. If the buffer of the root node now contains more than *m* blocks of such elements, perform a EMPTYBUFFER on the node as explained in Section 2.3.1 .

We will now state with proof, a simple theorem about insert operation.

**Theorem 1.1:** A new element is inserted in to the tree at an expected amortized cost of $O(\frac{\log_m n}{B})$ I/O's which is optimal.
*Proof*: The expected height of the random buffer tree is $O(\log_m n)$ as by Lemma 1. The EMPTYBUFFER primitive moves $\Theta(m)$ blocks of insert(or other) elements one level down the tree in an amortized cost of $\Theta(m)$ I/O's by Lemma 2.1 or one insert element is moved down one level of the tree in $O(1/B)$ I/O's amortized. The lower bound of $O(n \log_m n)$ on sorting [1] together with the bound on tree height implies that the cost of $O(\frac{\log_m n}{B})$ I/O's per insertion is optimal. The result follows.

*2.4.2 delete(key)*
Deletion is handled essentially in the same way as insertion.
The steps involved in *delete* are:
1. Construct a new (delete) element with the given key.
2. Set the priority of the element to zero.
3. When *B* such delete(or other) elements have been collected in internal memory, insert the block in the buffer of the root node.
4. If the buffer of the root node now contains more than *m* blocks of such elements, perform a EMPTYBUFFER on the node as explained in Section 2.3.1.

We will now state with proof, a simple theorem about delete operation.

**Theorem 1.2:** An existing element in the tree is deleted at an expected amortized cost of $O(\frac{\log_m n}{B})$ I/O's which is optimal.

*Proof*: The expected height of the random buffer tree is $O(\log_m n)$ by Lemma 1. The EMPTYBUFFER primitive moves $\Theta(m)$ blocks of delete(or other) elements one level down the tree in an amortized cost of $\Theta(m)$ I/O's by Lemma 2.1 or one delete element is moved down one level of the tree in $O(1/B)$ I/O's amortized. The lower bound of $O(n \log_m n)$ on sorting [1] together with the bound on tree height implies that the cost of $O(\frac{\log_m n}{B})$ I/O's per delete is optimal. The result follows.

*2.4.3 search(key)*
The search operation is used to lookup an existing element with the given *key*.
The steps involved in *search* are:
1. Construct a new (search) element with the given key and priority zero.
2. When $B$ such search(or other) elements have been collected in internal memory, insert the block in the buffer of the root node.
3. If the buffer of the root node now contains more than $m$ blocks of such elements, perform a EMPTYBUFFER on the node as explained in Section 2.3.1.

We will now state with proof, a simple theorem about search operation.

**Theorem 1.3:** An existing element in the tree is queried at an amortized cost of $O(\frac{\log_m n}{B})$ I/O's which is optimal.

*Proof*: The expected height of the random buffer tree is $O(\log_m n)$ by Lemma 1. The EMPTYBUFFER primitive moves $\Theta(m)$ blocks of search(or other) elements one level down the tree in an amortized cost of $\Theta(m)$ I/O's by Lemma 2.1 or one search element is moved down one level of the tree in $O(1/B)$ I/O's amortized. The lower bound of $O(n \log_m n)$ on sorting [1] together with the bound on tree height implies that the cost of $O(\frac{\log_m n}{B})$ I/O's per search is optimal. The result follows.

*2.4.4 deletemin*
The *deletemin* operation is used to return the element with the minimum *key* and delete it from the tree. The random buffer tree supporting the deletemin operation may be used as an I/O-efficient priority queue.

A search tree structure can normally be used to implement a priority queue because the smallest element is in the leftmost leaf. The same strategy cannot immediately be used on the random buffer tree since the smallest element is not necessarily stored in the leftmost node – smaller elements could reside in the buffers of the node on the leftmost root-leaf path. However if we empty the buffers completely (perform a full EMPTYBUFFER) on all nodes along the path from the root to the leftmost leaf using $O(m \log_m n)$ expected amortized I/O's and then delete the *m/4* blocks of smallest elements from the tree which are now stored in the leftmost leaf node, and store it in main memory we can answer the next *m/4* blocks of deletemin requests without performing any further I/O's. We have to then update the list of minimal elements as insertions and deletions are performed on the tree and this can be done in main memory without incurring any further I/O's.

In some applications, it may be preferable to use less than *m/4* blocks of internal memory for the external memory priority queue structure. We can make our priority queue work with $(1/4)m^{1/c}$ ($0 < c \leq 1$) blocks of internal memory simply by decreasing the fan-out and the size of the buffers to $\Theta(m^{1/c})$, as the expected height of the tree remains $O(\log_m n)$ even with this reduced fan-out.

We will now state with proof, a simple theorem about deletemin operation.

**Theorem 1.4:** The minimal element in the tree is deleted from the tree and returned to main memory at an expected amortized cost of $O(\frac{\log_m n}{B})$ I/O's which is optimal.

*Proof*: The expected height of the random buffer tree is $O(\log_m n)$ by Lemma 1. The expected amortized cost of emptying $O(\log_m n)$ buffers along the path from root to leftmost leaf is $O(m \log_m n)$ I/O's. The next *m/4* blocks of deletemin requests are satisfied without any further I/O. Hence the expected amortized I/O per deletemin request is $O(\frac{\log_m n}{B})$ I/O's. The lower bound of $O(n \log_m n)$ on sorting [1] together with the bound on tree height implies that the cost of $O(\frac{\log_m n}{B})$ I/O's per deletemin is optimal. The result follows.

**2.5 The complexity analysis of the random buffer tree**

In this section we give the I/O and space complexity analysis of the random buffer tree.

**2.5.1 The I/O Complexity of the random buffer tree**

The primitives EMPTYBUFFER and FILLBUFFER perform all the I/O operations, each one moving a collection of elements one level down or up the tree. These functions are I/O efficient — that is, when they move x elements up or down one level, they need $O(x)$ I/O's to do so.

We are now ready to state with proof our main theorem.

**Theorem 2:** *The total expected amortized cost of an arbitrary sequence of N intermixed insert, delete, search and deletemin operations performed on an initially empty random buffer tree is* $O(n \log_m n)$ *I/O operations.*

*Proof*: The random buffer tree performs insert, delete, search and deletemin operations on an element in an optimal expected amortized $O(\frac{\log_m n}{B})$ by Theorem 1.1, 1.2, 1.3 and 1.4 respectively. The result follows.

These performance bounds for the random buffer tree lead immediately to an optimal randomized sorting algorithm for *N* elements, which consists of just *N* insertions followed by *N* deletemins.

**Corollary 1** On a random buffer tree, *N* elements are sorted optimally using expected $O(n \log_m n)$ I/O's using the standard tree-sort.

### 2.5.2 Space Complexity of Random Buffer Tree

The number of internal nodes in the tree is expected to be $O(n/m)$ and each internal node has a buffer of size *M*. Each element, denoted by a <key,priority> pair, is stored at only one location in the tree. Thus the space complexity of the tree is $O(n)$.

### 2.6 The Random Buffer Tree Vs The Buffer Tree

In this section, we make a comparative study of the random buffer tree and the buffer tree and point out the pros and cons of each structure.

The buffer tree is the most ubiquitous external memory data structure for dynamic offline problems, with an ever-increasing number of applications. The random buffer tree has the same expected I/O-complexity for the insert, delete, deletemin and search operations as the buffer tree in the worst case.

The advantages of the random buffer tree over the buffer tree are:
- **Simpler tree re-balancing logic**: The random buffer tree is a self-balancing data structure based on the random *m*-ary search tree and the re-balancing routines upon insertion (EMPTYBUFFER) and deletion (FILLBUFFER) from the tree are very simple. The buffer tree, on the other hand, has a re-balancing logic, that is quite complicated by comparison.
- **Oblivious data structure:** An oblivious data structure [7] is one which does not reveal anything about its history; in particular about the order of insertions and deletions. A randomized search tree (like the random buffer tree) has this property, whereas an (a,b)-tree (like the buffer tree) does not. The random buffer tree provides a simple history-independent implementation of the dictionary data structure. Oblivious data structures have important cryptographic applications.
- **Easy to make self-adjusting:** It is relatively easy to convert the random buffer tree to a self-adjusting index suitable for applications with skewed access patterns [*see* section 2.7], whereas self-adjusting versions of the buffer tree are not known.

- **Ease of implementation:** It is expected that the random buffer tree, with its simpler specification, will provide a simpler alternative to the buffer tree in external memory applications.

The disadvantages of the random buffer tree over the buffer tree are:
- **Expected bounds:** The random buffer tree provides *expected* amortized bounds on the standard operations, whereas the buffer tree gives *exact* amortized bounds on the operations.
- **Space Overhead:** The random buffer tree requires more space compared to the buffer tree, as a priority value has to be stored for each element, in addition to the key value. However, the total space requirement is still linear in the number of elements stored in the tree.

**2.7 Self-Adjusting Random Buffer Tree**

In this section, we discuss two heuristics to transform the random buffer tree into a self-adjusting structure, suitable for applications with skewed access patterns.

In the first scheme due to Siedel et al. [8], in order to ensure that frequently accessed elements get promoted towards the root of the random buffer tree, the following steps are taken on an access to an element in the tree. When an element X is accessed, a new random number, $r$, is generated. If $r$ is greater than the current priority of X, $r$ becomes the new priority of X; otherwise, nothing further is done on this access. The readjustments to the tree (FILLBUFFER) to maintain heap order will promote the element closer to the root of the tree. After the element X has been accessed k times, its priority will be the maximum of k i.i.d. random variables. Thus the expected depth of the element X in the tree will be $O(\log_m(1/p))$, where p is the access frequency of x, i.e. $p = k / A$, with A being the total number of accesses to the tree.

In the second scheme due to Bitner [9], an element's priority is explicitly determined by counting the number of times the node has been accessed. When an element is inserted into the tree it's priority (access counter) is set to 1 and each subsequent access to the element increments this counter by 1. As the accesses to a node increase, the readjustments to maintain heap order will promote the element closer to the root of the tree. There are two potential problems with such a method of determining priorities. The first is that the counters may get so high that the tree loses its ability to adapt easily to changes in the access distribution and the second is that eventually (when dealing with large sets of data that are accessed regularly) the access counters will overflow. To overcome these problems, the counter is represented with a fixed number of bits and the counters are reset when the counter of the root node reaches some pre-specified upper bound. This scheme performs poorly for normal distribution, but is expected to perform better where the data exhibits low entropy (skewed distribution).

## 3. Conclusion

The random buffer tree is an I/O-efficient probabilistic self-balancing data structure that has the same expected I/O performance bounds as the buffer tree in the worst case and has the additional virtue of having much simpler specification compared to the buffer tree, leading to simpler implementation. It is expected that random buffer tree will be the data structure of choice for implementing dictionaries in external memory, as the random treap is the preferred implementation of the dictionary in internal memory, for e.g., in LEDA[10] , an efficient library of algorithms and data structures.